\documentclass{nic-series}
\usepackage{epsf,psfig,epsfig}

\begin{document}

\title{Monte Carlo Protein Folding: \\
      Simulations of Met-Enkephalin 
      with Solvent-Accessible Area Parameterizations}

\author{Hsiao-Ping Hsu$^1$, Bernd A. Berg$^2$, and Peter Grassberger$^1$ }

\institute{$^1$Complex Systems Research Group, 
           John-von-Neumann Institute for Computing\\
           Forschungszentrum J{\"u}lich, D-52425 J{\"u}lich, Germany\\
           $^2$Department of Physics, Florida State University,
           Tallahassee, FL 32306}

\maketitle

\begin{abstracts}
Treating realistically the ambient water is one of the main difficulties
in applying Monte Carlo methods to protein folding. The solvent-accessible 
area method, a popular method for treating water implicitly, is 
investigated by means of Metropolis simulations of the brain peptide 
Met-Enkephalin. For the phenomenological energy function ECEPP/2 nine 
atomic solvation parameter (ASP) sets are studied that had been proposed 
by previous authors. The simulations are compared with each other, with 
simulations with a distance dependent electrostatic permittivity $\epsilon
(r)$, and with vacuum simulations ($\epsilon =2$). Parallel tempering and 
a recently proposed biased Metropolis technique are employed and their 
performances are evaluated. The measured observables include energy and 
dihedral probability densities (pds), integrated autocorrelation times, and
acceptance rates. Two of the ASP sets turn out to be unsuitable for
these simulations. For all other sets, selected configurations are
minimized in search of the global energy minima. Unique minima are found for
the vacuum and the $\epsilon(r)$ system, but for none of the ASP models.
Other observables show a remarkable dependence on the ASPs. In particular,
autocorrelation times vary dramatically with the ASP parameters. 
Three ASP sets have much smaller autocorrelations at 300~K 
than the vacuum simulations, opening the possibility that simulations
can be speeded up vastly by judiciously chosing details of the force 
field.
\end{abstracts}

\section{Introduction}
 
Protein folding is considered as one of the grand challenges in 
mathematical biology. Actually, there are several different problems
related to protein folding. Strictly speaking, one should distinguish
between {\it fold prediction}, i.e. the mapping of the amino acid 
sequence onto the geometry of the native configuration, and understanding 
the {\it pathway} along which the folding proceeds. Another problem
is {\it inverse} fold prediction, i.e. finding an amino acid sequence
which will fold into a desired native configuration. 

At present there is no hope that any of this problems can be attacked 
{\it ab initio}, i.e. by solving the many-body Schr\"odinger equation.
Instead, one first constructs effective potentials (``force fields")
which then allow the dynamics of the nuclei to be treated by 
classical mechanics. The native state is then identified with the 
state of lowest energy. There are several such force fields in current 
use, each one with its weaknesses and strengths. It seems fair to say 
that their precision typically is sufficient for the correct folding of 
peptides and small proteins, but not for larger ones with, say, more 
than 50 amino acids. In addition, some force fields make additional
simplifications such as keeping bond lengths fixed or even lumping 
several small atoms into one effective particle.

Given an amino acid sequence and a force field, finding the native 
state would then seem straightforward, if there were a single local
energy minimum. But, alas, energy landscapes for typical proteins are 
rough, with many local minima, and finding the global one is highly
non-trivial. Basically, there are two methods available for this 
purpose: molecular dynamics and Monte Carlo methods.

In molecular dynamics (MD) one just solves numerically Newton's 
equations of motion. This has the advantage that one simulates 
directly the physical folding process, i.e. one obtains directly 
the dominant folding paths and one gets immediately estimates for 
the folding times. There are of course many details which can be 
adjusted to make the simulation faster and more realistic
(e.g., one can include thermal noise and solve a Langevin equation,
or one can treat Coulomb forces more efficiently), but basically 
one has not much freedom in putting up the simulations, and these 
simulations tend to be slow -- very slow!

This is in contrast to Monte Carlo (MC) simulations. There one gives up
any claim to follow the actual path, but one jumps in phase space as 
efficiently as possible, subject only to the constraint that one 
samples configurations according to the Boltzmann-Gibbs distribution,
$p({\rm config}) \propto \exp(-E({\rm config})/k_BT)$. Indeed, with modern
advanced MC methods one may even give up this requirement and re-weight
eventually the final distribution to obtain proper sampling 
\cite{ls87,okk92,ho93,mmm94,hoo99,Okamoto,wille-hansmann,wang-landau,b03}.
Since one is completely free in how one moves in phase space (one may
even move by adding or removing particles, see \cite{grass}), such a 
strategy can be very efficient if the moves are well chosen -- but it 
can also be extremely inefficient, if they are badly chosen.

In nature biomolecules exist in the environment of solvents (i.e. water,
in general), thus the molecule-solvent interactions must be taken into 
account. Indeed, neglecting the ambient water altogether can lead to 
gross errors. In MD simulations water just slows down the simulations 
because the number of particles increases by a factor between two and 
ten. For MC simulations the situation is much worse. For steric reasons, 
many moves which would be very efficient {\it in vacuo}, become 
inefficient (i.e., are accepted with small probabilities) if the molecule
is surrounded by water. This is the main reason why chemists in general
prefer MD over MC methods.

Since it is so very computer time consuming to simulate proteins when
the surrounding water is treated explicitly, a number of approximate 
treatments of solvent effects have been developed, where the water is 
treated only {\it implicitly}. In the {\it solvent-accessible 
area approach}~\cite{lr71,c74,em86} it is assumed that the 
protein-solvent interaction is given by the sum of the surface area of 
each atomic group times a parameter called {\it atomic solvation parameter}
(ASP). The choice of a set of ASPs (also called hydrophobicity parameters or
simply hydrophobicities) defines a model of solvation.  However, there 
is no agreement on how to determine the universally best set of ASPs, 
or at least the best set for some limited purpose. For instance, eight 
sets were reviewed and studied by Juffer et al.~\cite{jeh95} and it was 
found that they give rather distinct contributions to the free energy.

In Ref.\cite{bh03} we investigated how different ASP sets modify the
Metropolis simulations of the small brain peptide Met-Enkephalin 
(Tyr-Gly-Gly-Phe-Met) at 300~K. The reason for the choice of 
Met-Enkephalin is that its vacuum properties define a reference 
system for testing numerical methods, 
e.g.~\cite{ls87,okk92,ho93,mmm94,hoo99,b03}. 
Therefore, Met-Enkephalin appears to be well suited to set 
references for the inclusion of solvent effects as well, but we 
are only aware of few articles~\cite{ls88,koh98}, 
which comment on the modifications due to including a solvent model. 
On the other hand, the effect of ASP models on the helix-coil transition 
of polyalanine has been studied recently~\cite{PeHa03}.

The simulation temperature was chosen as 300~K in Ref.\cite{bh03}, because 
room temperature is the physical temperature at which biological activity 
takes place. Most of the previous simulations of Met-Enkephalin in vacuum 
were performed at much lower temperatures or employed elaborate 
minimization techniques with the aim to determine the global energy 
minimum (GEM). Only recently~\cite{b03} it was shown that the GEM is 
well accessible by local minimization of properly selected configurations 
from an equilibrium time series at 300~K. Precisely this should be the 
case for a GEM which is of relevance at physical temperatures. 

For our simulations we use the program package SMMP~\cite{smmp} 
(Simple Molecular Mechanics for Proteins) together with parallel
tempering~\cite{g91,hn96,ha97} (PT) and the recently introduced~\cite{b03} 
biased Metropolis technique RM$_1$ (rugged Metropolis -- 
approximation~1). SMMP implements a number of all-atom energy 
functions, describing the intramolecular interactions, and nine 
ASP sets~\cite{em86,oons,sch1,sch2,jrf,we92,sch4,bm} to model the
molecule solvent interactions. We use the ECEPP/2~\cite{sns84} 
(Empirical Conformational Energy Program for Peptides) energy 
function with fully variable $\omega$ angles and
simulate all nine ASP sets. For comparison we simulate also 
Met-Enkephalin in vacuum and with the distance dependent 
electrostatic permittivity $\epsilon (r)$ of Ref.~\cite{hrf85}.

The energy functions and some details of the Metropolis methods used are 
explained in Sec.~\ref{basics}. In Sec.~\ref{results} we present the main 
results. Summary and conclusions are given in Sec.~\ref{conclusions}.

\section{Models and Methods} \label{basics}

\subsection{Force field and atomic solvation parameter sets}

In all-atom models of biomolecules the total conformational energy of 
the intramolecular interactions $E_I$ is given as the sum of the 
electrostatic, the Lennard-Jones (Van der Waals), the hydrogen bond, 
and the torsional contributions,
\begin{eqnarray}
E_I\ =\ 332\sum_{i<j} \frac{q_i q_j}{\epsilon r_{ij}}\ +\ \sum_{i<j} 
  \left(\frac{A^{\rm LJ}_{ij}}{r_{ij}^{12}}  \nonumber  
- \frac{B^{\rm LJ}_{ij}}{r_{ij}^6}\right)\ +\\ 
  \sum_{i<j} \left(\frac{A^{\rm HB}_{ij}}{r_{ij}^{12}}
- \frac{B^{\rm HB}_{ij}}{r_{ij}^{10}}\right)\ +\ 
\sum_k U_k\, [1\pm \cos(n_k \phi_k)] \, . \label{eceep2}
\end{eqnarray}
Here $r_{ij}$ is the distance between atoms $i$ and $j$, $q_i$ and $q_j$ 
are the partial charges on the atoms $i$ and $j$, $\epsilon$ is the 
electric permittivity of the environment, $A_{ij}$, $B_{ij}$, $C_{ij}$ 
and $D_{ij}$ are parameters that define the well depth and width for a 
given Lennard-Jones or hydrogen bond interaction, and $\phi_k$ is the 
$k$-th torsion angle. The units are as follows: distances are in \AA,
charges are in units of the electronic charge and energies are in
kcal/mol.

One of the simplest ways to include interactions with water is to
assume a distance dependent electrostatic permittivity according to
the formula~\cite{hrf85,ok94}
\be \label{epsilon}
\epsilon(r) = D-\frac{D-2}{2}\,\left[ (sr)^2+2sr+2\right]\,e^{-sr}\ .
\ee
Empirical values for the parameters $D$ and $s$ are chosen so that
the permittivity takes the value of bulk water, $\epsilon =80$,
for large distances, and the value $\epsilon=2$ for small $r$, i.e.
for the interior of the molecule. This approach is clearly an 
oversimplification, because atoms which are close to each other do not 
necessarily have to be simultaneously in the interior of the molecule.
Reversely, two atoms which are separated by a large distance may still 
be in the interior of the molecule. More elaborated approaches are
asked for.

   If the molecule-solvent interaction is proportional to the surface 
area of the atomic groups, it is given by the sum of contributions of 
a product of the surface area of each atomic group and the atomic
solvation parameter~\cite{em86},
\be \label{Esol}
    E_{\rm sol}=\sum_i \sigma_i A_i \, .
\ee
Here $E_{\rm sol}$ is the solvation energy and the sum is over all
atomic groups.  $A_i$ is the solvent accessible surface area and 
$\sigma_i$ the atomic solvation parameter of group $i$. The choice 
of a set of ASPs $\sigma_i$ defines a model of solvation. In our work,  
we used the same nine sets of ASPs as in the SMMP package and refer 
to \cite{smmp,bh03} for notations and details.

\subsection{Metropolis methods} \label{sec_met}

For the updating we used PT with two processors, one 
running at 300~K and the other at 400~K. This builds on the 
experience~\cite{b03} with vacuum simulations of Met-Enkephalin for 
which the following observations are made:

\begin{enumerate}
\item The integrated autocorrelation time $\tau_{\rm int}$ (defined
below in this section) increases from 400~K to 300~K by a factor
of ten for the (internal) energy and by factors of more than twenty for
certain dihedral angles.
\item The energy probability densities (pds) at 300~K and 400~K 
overlap sufficiently, so that the PT method works, leading to an 
improvement factor of about 2.5 in the real time needed for the 
simulation (see Table~I of Ref.~\cite{b03}).
\end{enumerate}

A detailed description of the PT algorithm is given in Ref.\cite{bh03}.
It used an approximation, called RM$_1$, to the rugged Metropolis
scheme introduced in Ref.\cite{b03} which had given an improvement
by an additional factor of two for the vacuum system~\cite{b03}.
Finally, the GEM was determined by minimizing selected configurations
of the 300~K time series. More precisely, 

\begin{enumerate}
\item We determined the lower 10\% quantile $E_{0.1}$ and the upper
10\% quantile $E_{0.9}$ of the energy distribution of our time 
series. This is done by sorting all energies in increasing order
and finding the values which cut out the lower and upper 10\% of
the data. For the statistical concepts see, e.g., Ref.~\cite{brandt}.

\item We partitioned the time series into bunches of configurations.
A bunch contains the configurations from one crossing of the
upper quantile $E_{0.9}$ to the next so that at least on crossing
of the lower quantile $E_{0.1}$ is located between the two crossings
of $E_{0.9}$. For each bunch we picked then its configuration of 
lowest energy. The idea behind this procedure is to pick minima of 
the time series, which are to a large degree statistically independent.

\item We run a conjugate gradient minimizer on all the selected
configurations and thus obtain a set of configurations which are 
local energy minima. For the vacuum simulation~\cite{b03} about
5\% to 6\% of the thus minimized configurations agreed with the
GEM.
\end{enumerate}

To determine the speed at which the systems equilibrate, we measured the
integrated autocorrelation time $\tau_{\rm int}$ for the energy and each 
dihedral angle. These times are 
directly proportional to the computer run times needed to achieve the
same statistical accuracy for each system. They thus determine the 
relative performance of distinct algorithms. For an observable $f$ the 
autocorrelations are 
\be
 C(t) = \langle f_0\,f_t\rangle - \langle f\rangle^2
\ee
where $t$ labels the computer time. Defining $c(t)=C(t)/C(0)$, the
time-dependent integrated autocorrelation time is given by
\be \label{tau_int_t}
 \tau_{\rm int}(t) = 1 + 2 \sum_{t'=1}^t c(t')\ .
\ee
Formally the integrated autocorrelation time $\tau_{\rm int}$ is 
defined by $\tau_{\rm int}=\lim_{t\to\infty}\tau_{\rm int}(t)$.
Numerically, however, this limit cannot be reached as the noise 
of the estimator increases faster than the signal. Nevertheless,
one can calculate reliable estimates by reaching a window of $t$ 
values for which $\tau_{\rm int}(t)$ becomes flat, while its error 
bars are still reasonably small. The data given in the next section 
were obtained in this way, see Ref.~\cite{sokal} for a more detailed 
discussion.

\section{Results} \label{results}

\subsection{Autocorrelations}

The PT simulations with temperatures $T_0=400\,$K and $T_1=300\,$K are 
performed on the system in vacuum ($\epsilon=2$), with $\epsilon(r)$ 
given by Eq.~(\ref{epsilon}) and for the nine ASP sets in the SMMP
package. The ranges of the dihedral angles are not restricted but 
vary in the full range from $-\pi$ to $\pi$. Each measurement is 
based on $\approx 2\times 10^6$ sweeps, where one sweep is defined by 
updating each dihedral angle once. On the Cray T3E, this takes about 
14 hours for the vacuum system and $5\times 14$ hours for each ASP set.

\begin{figure}[ht]
\begin{center}
$\begin{array}{c@{\hspace{0.1in}}c}
\multicolumn{1}{l}{\mbox{\small (a)}} &
        \multicolumn{1}{l}{\mbox{\small (b)}} \\ [-0.53cm]\\
\epsfig{file=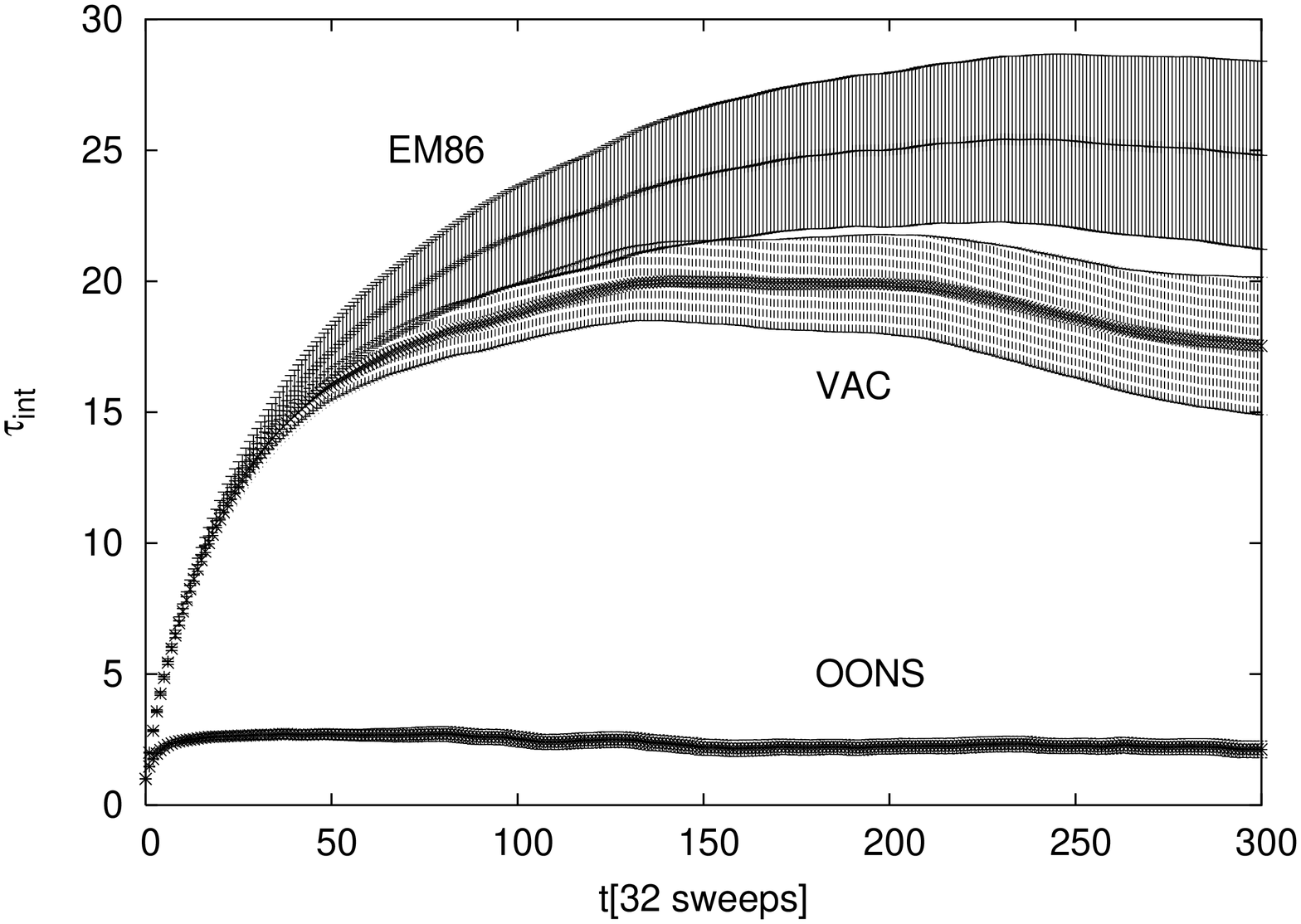, width=4.3cm, angle=270} &
\epsfig{file=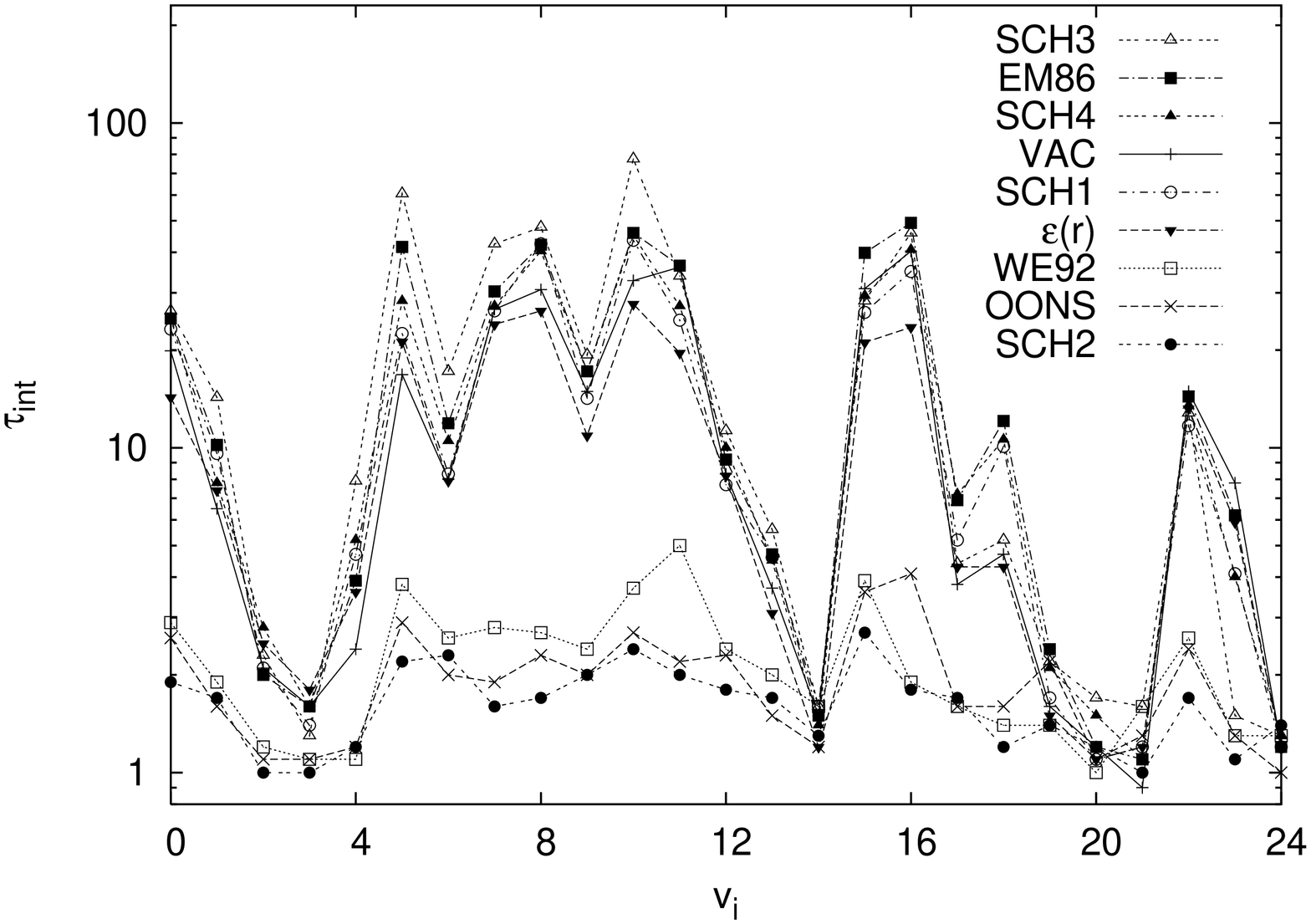, width=4.3cm, angle=270} \\
[0.4cm]
\end{array}$
\caption{
(a) The time-dependent integrated autocorrelation time
   for the energy at 300~K from our simulations of the vacuum
   system and the solvent models EM86 and OONS.
(b) Integrated autocorrelation times for the energies
   ($v_0=E$) and the dihedral angles $v_i$, $i=1,\dots,24$ at
    $T=300\,$K.}
\label{fig_taug}
\end{center}
\end{figure}

Results of the time-dependent integrated autocorrelations
times~(\ref{tau_int_t}) for the vacuum simulations and the ASP sets
OONS and EM86 are shown in Fig.~\ref{fig_taug}a. 
In
each case a window of $t$ values is reached for which
$\tau_{\rm int}(t)$ does no longer increase within its statistical
errors. In the case of the vacuum simulations it even decreases,
but this is not significant due to the statistical error. These
windows are then used to estimate the asymptotic $\tau_{\rm int}$
for all ASP sets except the ASP sets JRF and BM. 

The acceptance rates of the solvent 
models JRF and BM are much lower than for the other models. In essence 
the simulations of these two models get stuck, which implies that their 
integrated autocorrelation times cannot be measured. 
The pds of the dihedral angles
of these two models are also erratic and the conclusion is that they
cannot be used to describe Met-Enkephalin in solvent.

The energy couples to all dihedral angles and its integrated 
autocorrelation time is characteristic for the entire system, while
the integrated autocorrelation times of the single dihedral angles 
vary heavily from angle to angle. For all systems but JRF and BM,
we show in Fig.~\ref{fig_taug}b the integrated autocorrelation times 
at 300~K for the energy and all dihedral angles. The notation 
$v_i$, $i=0,1,\dots,24$ is used, where $v_0$ is stands in for the
energy $E$ and for $i=1,\dots ,24$.

In Fig.~\ref{fig_taug}b we see that for each dihedral angle $v_i$ 
the integrated autocorrelation times $\tau_{\rm int}[v_i]$ for the 
three solvent models OONS, WE92 and SCH2 are smaller than for the 
remaining systems, including the vacuum system. 
In particular, this means that the OONS, WE92 and SCH2 models require 
far less statistics than the vacuum run for achieving the same 
accuracy of results. In the following
the solvation models OONS, WE92 and SCH2 define the ``fast class'',
while the other models shown in Fig.~\ref{fig_taug}b constitute
the ``slow class'' (the models JRF and BM are omitted from this
classification). ``Good'' behavior of the models OONS and WE92
has been previously observed~\cite{ok98}.
Precise values of the autocorrelation times and further details on 
their measurements can be found in \cite{bh03}.

\subsection{Structure}

For all our simulations we applied the method outlined in 
subsection~\ref{sec_met} to determine local energy minima. Again, the 
results of the JRF and BM solvent model are erratic. The BM model is 
entirely frozen, only $N_{\rm conf}=2$ different configurations are 
ever reached at 400~K and $N_{\rm conf}=1$ at 300~K. Therefore, 
we do not give minimization results for BM. For JRF the 
$N_{\rm conf}$ numbers are more reasonable, but still by a factor of 
one third and less smaller than the $N_{\rm conf}$ numbers of each 
other system. JRF is also disregarded in the following discussion.

Only if the same energy minimum is hit $N_{\rm hits}>1$ times, we
can argue that we found the GEM. This was the case for the vacuum and 
for the $\epsilon(r)$ simulations (notably already at 400~K), 
but not for any of the ASP solvent models. There, each minimization
led to a different state. This is a very interesting observation,
as it might indicate that the energy landscape is rougher for 
the ASP solvent models. But if this were the case, we should also
expect that autocorrelations are longer for the ASP solvent models,
while the opposite was found at least for three ASP parameter sets.

Indeed we were not the first to observe this phenomenon.
Quite some time ago Li and Scheraga~\cite{ls87,ls88} 
developed a Monte Carlo minimization method and applied it to 
Met-Enkephalin in vacuum and in solvent modeled by OONS. While 
for the vacuum system their method converged consistently to the 
GEM, all their five runs of the solvent model led to different 
conformations with comparable energies. They interpreted their 
results in the sense that Met-Enkephalin in water at $20^{\circ}$C 
is presumably in an unfolded state for which a large ensemble of 
distinct conformations co-exist in equilibrium. A consistent
scenario was later observed in NMR experiments~\cite{grcahi}.

Although the minimization method of  Li and Scheraga is entirely 
different from ours, they essentially tested for valleys of attraction 
to the GEM at room temperature, quite as we do in the present 
paper. So, we have not only confirmed their old result, but find 
that it is common to a large set of ASP models implemented in 
SMMP. Neither the method by which an ASP set was derived, nor 
whether it belongs to the fast or slow class, appears to matter 
with this respect. 

\begin{figure}[ht]
\begin{center}
$\begin{array}{c@{\hspace{0.1in}}c}
\multicolumn{1}{l}{\mbox{\small (a)}} &
        \multicolumn{1}{l}{\mbox{\small (b)}} \\ [-0.53cm]\\
\epsfig{file=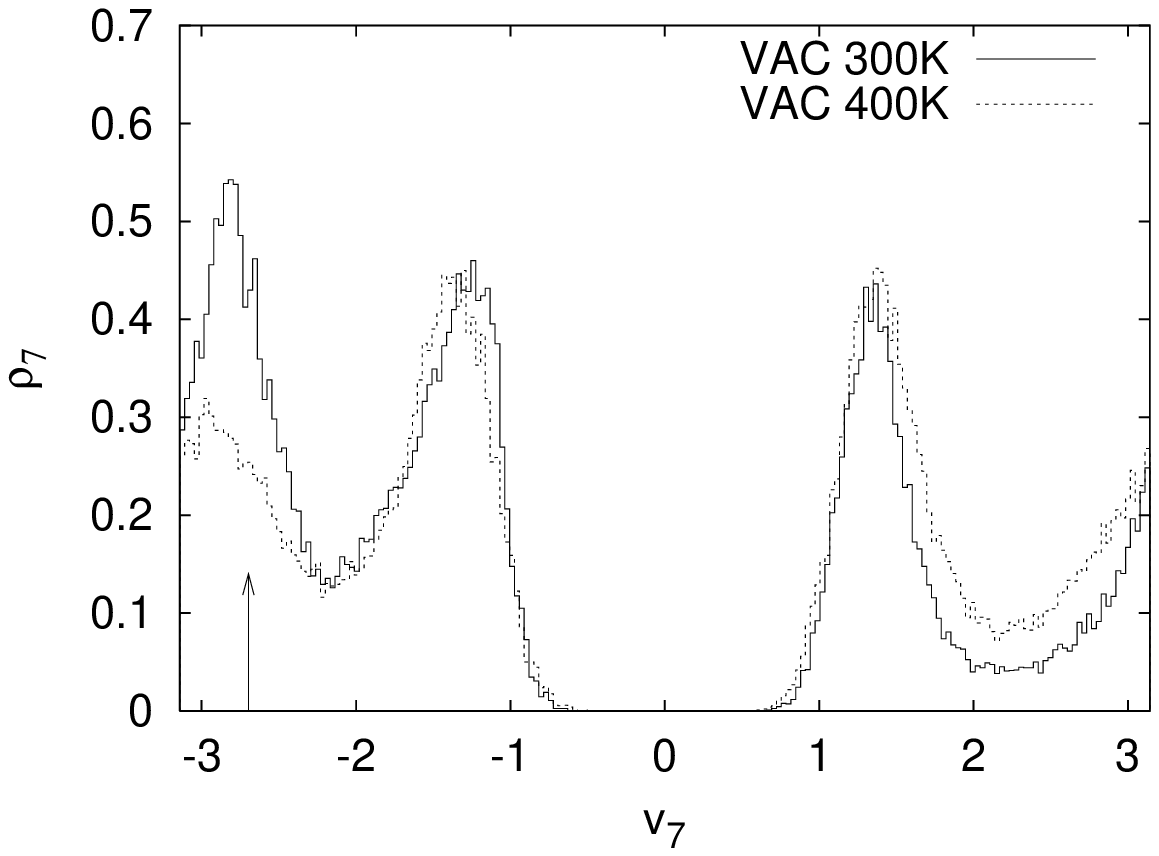, width=6.14cm, angle=0} &
\epsfig{file=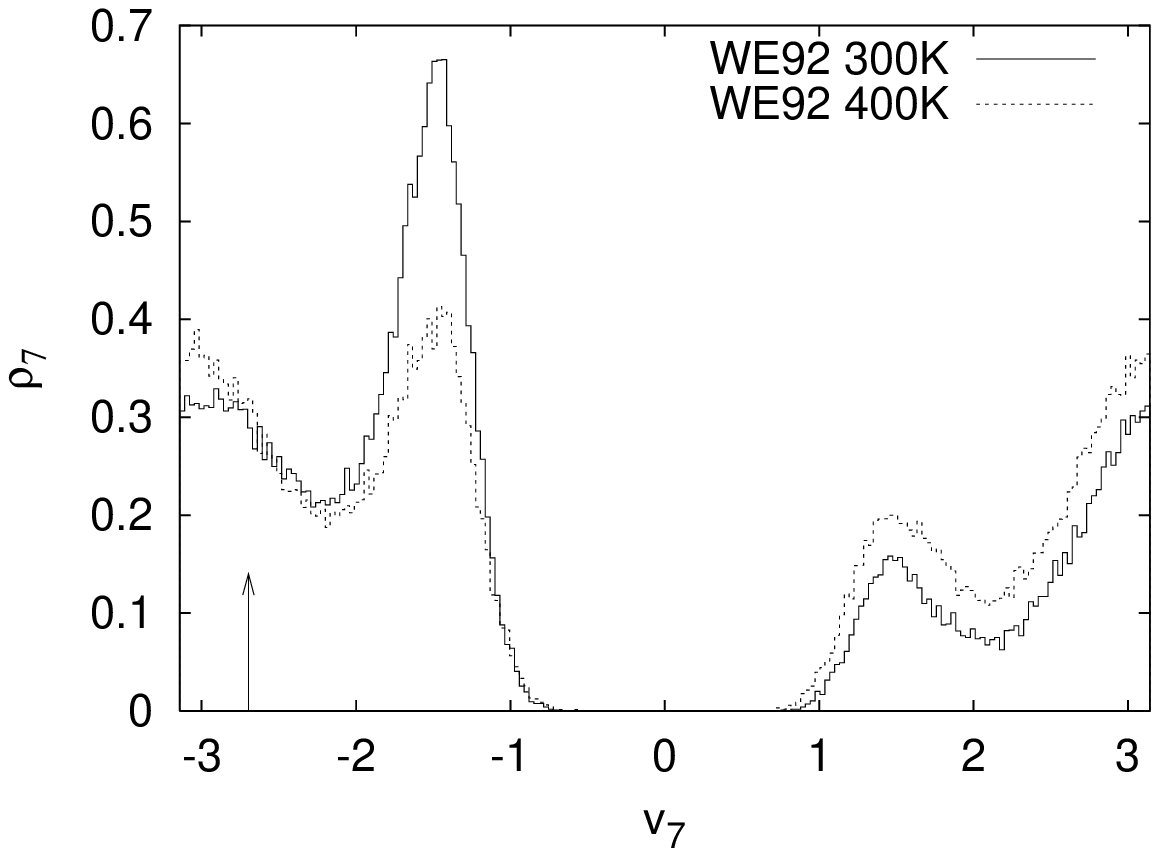, width=6.14cm, angle=0} \\
[0.4cm]
\end{array}$
\caption{
Probability densities of the dihedral angle $v_7$ for (a)
    the vacuum simulation and (b) the WE92 simulation.
    The arrow indicates the vacuum GEM value of this angle.}
\label{fig_v7}
\end{center}
\end{figure}

In a search for structural differences of Met-Enkephalin in
the different models, we looked at the pds of the dihedral angles.
For all systems and both temperatures there are altogether 
$2\times 9\times 24 = 432$ distributions. At the first look 
the pds of the different
systems are amazingly similar, independently of whether they are 
from systems of the fast or slow class, from an ASP model, from the 
vacuum or from the $\epsilon(r)$ simulation.
A more careful investigation reveals differences which appear 
to relate to the distinct behavior under our minimization. For the 
dihedral angle $v_7$ this is illustrated in Fig.~\ref{fig_v7}a and
Fig.~\ref{fig_v7}b. Its probability densities are compared at 
300~K and  400~K. For the vacuum simulation the pds are 
depicted in Fig.~\ref{fig_v7}a and from 400~K to 300~K 
we observe an increase of the peak close to the arrow which 
indicates the vacuum GEM value of $v_7$. In contrast to this, 
the wrong peak increases for the WE92 solvent model 
(Fig.~\ref{fig_v7}b).

\begin{figure}
  \begin{center}
\psfig{file=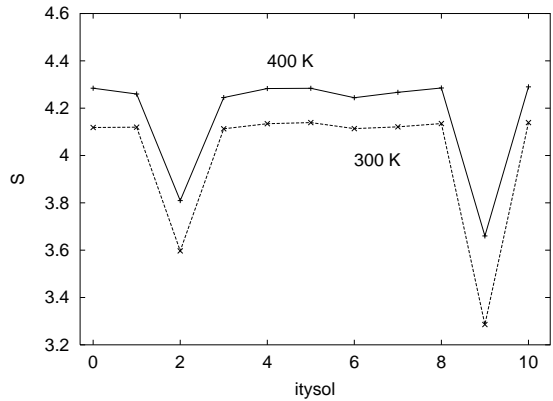,width=5.4cm, angle=270}
   \caption{Overall entropies of the dihedral angles.
   The numbers on the $x$-axis label the different models, with 
   ${\tt itysol}=0$ for vacuum and ${\tt itysol}=10$
   for the $\epsilon(r)$ model.} \label{fig_entropy}
  \end{center}
\end{figure}

One may suspect that the difference between the models of our fast
and slow classes is simply due to an effectively higher temperature
for the three models of the fast class. To gain insight into this
question, we calculate the corresponding entropies. Each pd is 
discretized as a histogram of 200 bins, $\rho_{ij}$, where $i=1,
\dots,24$ labels the dihedral angles, and $\sum_{j=1}^{200}\rho_{ij}=1$. 
The entropy of a dihedral angle is then defined by
\be \label{entropy}
S_i = - \sum_{j=1}^{200} \rho_{ij}\,\ln \rho_{ij}
\ee
and the total entropy of the pds of an ASP model is $S=\sum_iS_i$.
In Fig.\ref{fig_entropy} these entropies are depicted
for all our models. The lines between the data points are just 
drawn to guide the eyes. The dips for the JRF and the BM model 
show, again, that their configurations are essentially frozen. For 
the other models we see a decrease of entropy from 400~K to 300~K,
but we find no larger entropy for the models of the fast class than
for the models of the slow class. Therefore, the effective 
temperature scenario is ruled out. Instead, it seems that for
the models of the fast class the solvent has some kind of 
``lubrication'' effect, which accelerates the simulation.

Strong similarities between the ASP models of the fast class on one side
and the ASP models of the slow class on the other side are found for the
solvation energies, the gyration radii and the end-to-end distances.

\section{Summary and Conclusions} \label{conclusions}

We have reported on simulations of Met-Enkephalin at room temperature 
(300~K) for nine different solvation models. 
Quantitative results obtained in that way should not be trusted, 
because the methods to derive the ASPs have been quite 
crude. Also our simulations do not give information that would 
allow us to pick a best ASP set for the intended purpose of 
simulating Met-Enkephalin at 300~K. Nevertheless, we obtained
a number of very interesting consequences which should be more
general and which should apply to any attempt to include
solvation effects in Monte Carlo calculations.

If we exclude two ASP sets which behave erratic (at least as 
implemented in SMMP~\cite{smmp}), we have still nine models:
seven ASP sets, vacuum simulations with $\epsilon=2$, and the 
$\epsilon(r)$ system~\cite{hrf85}. These models seperate into 
a fast and a slow class with respect to their autocorrelation times. 
Vacuum simulations are in the slow class. This leads to the 
interesting feature that it takes less computer time to estimate 
physical observables at room temperature in the fast solvation 
models OONS~\cite{oons}, WE92~\cite{we92}, and 
SCH2~\cite{sch2,sch4} than it takes for vacuum, despite the 
substantial increase of the computer time per sweep by a factor 
of about 5 for the solvation models over the vacuum system. We 
have no clear clue why some models have a fast and others a slow 
dynamics. But the possibility that slightly different 
force fields can lead to vastly different autocorrelation times
is of course something which should be kept in mind.

We applied the minimization procedure of Ref.\cite{b03} in an
attempt to locate the GEM for the nine systems which are 
reasonably well-behaved under Metropolis simulations at 300~K.
The GEM is unambiguously found for the vacuum system and for
the simulation with a distance dependent electrostatic permittivity.
No true GEM is found for any of the remaining seven ASP models. This 
confirms an old result of Li and Scheraga~\cite{ls88}, who 
concluded that at room temperature Met-Enkephalin in water is 
likely in an unfolded state.
To get a better understanding of this result, we studied at
300~K the dihedral pds in some details. At a first glance they
look quite similar for all the models in the fast as well as in the 
slow class. Differences are found for a number of details, which
may allow to explain why the 300~K configurations of the ASP 
models behave entirely different under our minimization procedure 
than the vacuum and the $\epsilon(r)$ systems. 

The central question, which remains to be settled, is whether ASP 
models will ultimately allow for accurate Metropolis simulations 
of biomolecules like Met-Enkephalin in solvent or not. In principle,
this could be decided by comparing with simulations with 
explicit solvent, but this has not yet been done.

\bigskip

The computer simulations were
carried out on the Cray T3E of the John von Neumann Institute for
Computing. BB acknowledges partial support by the U.S. Department 
of Energy under contract No. DE-FG02-97ER41022.

\end{document}